\documentclass{aa}

\usepackage{graphicx}
\usepackage[table]{xcolor}
\usepackage{txfonts}

\begin{document}

   \title{SVOM discovery of a strong X-ray outburst of the blazar 1ES~1959+650 and multi-wavelength follow-up with the Neil Gehrels Swift observatory}

          \author{A. Foisseau
          \inst{1}\thanks{Corresponding author: \email{foisseau@apc.in2p3.fr}}
          \and
          A. Coleiro 
          \inst{1}
          \and
          S. Komossa 
          \inst{2,6}
          \and
          D. Grupe 
          \inst{3}
          \and
          F. Cangemi 
          \inst{1}
          \and
          P. Maggi\inst{4}
          \and
          D. Götz\inst{5}
          \\[1ex] 
          H.-B. Cai\inst{6} \and B. Cordier\inst{5} \and N. Dagoneau\inst{5} \and Z.-G. Dai\inst{7} \and Y.-W. Dong\inst{8} \and M. Fernandes Moita\inst{5} \and O. Godet\inst{9} \and A. Goldwurm\inst{1,10} \and H. Goto\inst{5} \and S. Guillot\inst{9} \and L. Huang\inst{6} \and M.-H. Huang\inst{6} \and N. Jiang\inst{7} \and C. Lachaud\inst{1} \and S. Le Stum\inst{1} \and E.-W. Liang\inst{11} \and X.-M. Lu\inst{6} \and L. Michel\inst{4} \and C. Plasse\inst{5} \and Y.L. Qiu\inst{6} \and J. Rodriguez\inst{5} \and L. Tao\inst{8} \and S. Schanne\inst{5} \and J. Wang\inst{6} \and X.-G. Wang\inst{11} \and X.-Y. Wang\inst{12} \and J. Wei\inst{6, 13} \and C. Wu\inst{6} \and Y.-W. Yu\inst{14} \and J. Zhang\inst{14} \and L. Zhang\inst{8}  \and S.-N. Zhang\inst{8} \and S. Zheng\inst{8}
         }

   \institute{
Université Paris Cité, CNRS, Astroparticule et Cosmologie, F-75013 Paris, France
\and
            Max-Planck-Institut für Radiostronomie, Auf dem Hügel 69, 53121 Bonn, Germany 
         \and 
             Department of Physics, Geology, and Engineering Technology, Northern Kentucky University, 1 Nunn Drive, Highland Heights, KY 41099, USA
        \and
            Observatoire Astronomique de Strasbourg, Université de Strasbourg, CNRS, 11 rue de l’Université, F-67000 Strasbourg, France
        \and
            Université Paris-Saclay, Université Paris Cité, CEA, CNRS, AIM, 91191, Gif-sur-Yvette, France
        \and
            National Astronomical Observatories, Chinese Academy of Sciences, Beijing 100101, People’s Republic of China
        \and
            Department of Astronomy, University of Science and Technology of China, Hefei 230026, People’s Republic of China
      \and
            Key Laboratory of Particle Astrophysics, Institute of High Energy Physics, Chinese Academy of Sciences, Beijing 100049, China
        \and 
            IRAP, CNRS, 9 avenue du Colonel Roche, BP 44346, F-31028 Toulouse Cedex 4, France
        \and
            CEA Paris-Saclay, Irfu / D´epartement d’Astrophysique, F-91191 Gif-sur-Yvette, France
        \and
            Guangxi Key Laboratory for Relativistic Astrophysics, School of Physical Science and Technology, Guangxi University, Nanning 530004,
People’s Republic of China
        \and 
            School of Astronomy and Space Science, Nanjing University, 210023 Nanjing, Jiangsu, China
        \and
            School of Astronomy and Space Science, University of Chinese Academy of Sciences, Beijing, People’s Republic of China
        \and
            School of Physics, Beijing Institute of Technology, Beijing 100081, People’s Republic of China
        }

   \authorrunning{Foisseau et al.}
   \titlerunning{SVOM discovery of a new X-ray outburst of the blazar 1ES~1959+650}

 \abstract
{On December 6, 2024, 1ES~1959+650, one of the X-ray brightest blazars known, underwent a high-amplitude X-ray outburst detected by SVOM, the first such discovery with this mission. The source was subsequently monitored with SVOM and \textit{Swift} from December 2024 to March 2025.}
{We report the detection and multi-wavelength follow-up of this event, and describe the temporal and spectral evolution observed during the campaign.}
{Data from SVOM/MXT, SVOM/ECLAIRs, and \textit{Swift}/XRT were analyzed with log-parabola models to track flux and spectral variability.}
{The source was detected in a bright state over the 0.3–50 keV range. During the three months of monitoring, the X-ray flux varied significantly, showing episodes of spectral hardening at high flux levels. The spectral curvature evolved more irregularly and did not show a clear trend with flux. A shift of the Spectral Energy Distribution (SED) synchrotron peak to higher energies is seen when the flux increases.}
{This constitutes the first blazar outburst discovered in X-rays by SVOM. The coordinated follow-up with \textit{Swift} provided continuous coverage of the flare and highlights the strong complementarity of the two missions for time-domain studies of blazars. The flare shows no clear signatures of either Fermi I or Fermi II acceleration, suggesting a mixed Fermi I/II scenario. }

   \maketitle

\section{Introduction}

Blazars are among the most extreme active galactic nuclei (AGNs), characterized by relativistic jets oriented close to the line of sight, producing strong Doppler boosting and large-amplitude variability across the electromagnetic spectrum \citep[see, e.g.,][]{1995PASP..107..803U}. Their broad-band spectra are non-thermal, with the low-energy component widely attributed to synchrotron emission from relativistic electrons, while the origin of the high-energy component is still debated between leptonic models and hadronic scenarios \citep{1985ApJ...298..114M, 1992A&A...256L..27D, 2013ApJ...768...54B}. The extreme conditions in blazar jets make them bright and highly variable sources, especially at X-rays and $\gamma$-rays energies.

1ES~1959+650 (z = 0.0048; \citealt{2004ApJ...601..151K}) is a high-synchrotron-peaked BL Lac object (HBL), with its synchrotron peak located in the Ultraviolet (UV) –X-ray range. The source was first detected in the radio and X-ray domains \citep{1991ApJS...75.1011G, 1992ApJS...80..257E}, and later in the very-high-energy (VHE) $\gamma$-ray band \citep{1999ICRC....3..370N} with the Whipple Observatory and the High-Energy-Gamma-Ray Astronomy (HEGRA) telescope. A major flaring episode was observed in 2002 at VHE $\gamma$-rays \citep{2003ApJ...583L...9H, 2003A&A...406L...9A, 2005ApJ...621..181D}, and since then, recurrent variability has been reported in optical \citep{2016ATel.9070....1B}, X-rays \citep{2016MNRAS.457..704K, 2018Galax...6..125K}, and $\gamma$-rays \citep[see, e.g.,][]{2016ATel.9010....1B, 2016ATel.9139....1B}. The source has also been studied as a potential emitter of high-energy neutrinos \citep{2005APh....23..537H} but it remains undetected. 

On December 6, 2024, the Chinese–French mission SVOM (Space Variable Objects Monitor; \citealt{2016arXiv161006892W}) detected 1ES~1959+650 in a bright X-ray flaring state \citep{2024ATel16935....1C}, marking the first blazar outburst discovered in X-rays by SVOM. A dense follow-up campaign was carried out in coordination with the Neil Gehrels Swift Observatory (hereafter \textit{Swift}; \citealt{2004ApJ...611.1005G}), providing optical to X-ray coverage from December 2024 to March 2025. In this article, we report the results of this campaign and discuss the spectral and temporal properties of the flare.

Section~2 describes the detection of the outburst phase as well as the monitoring campaign with both SVOM and \textit{Swift}. Section~3 presents the results, first describing the outburst evolution and then the spectral analysis. Section~4 is dedicated to the discussion and conclusions.

\section{Observations and data analysis}

    On 2024 December 6 at 15:08:48~UT, during its commissioning phase, the SVOM/ECLAIRs telescope (4--150~keV) detected enhanced X-ray emission from the field of the BL~Lac blazar 1ES~1959+650 \citep{2024ATel16935....1C} through the onboard trigger system. The source appeared as a previously unreported transient in ECLAIRs data, prompting rapid follow-up. A Target of Opportunity (ToO) observation was performed with SVOM, allowing MXT (0.2--10~keV) and VT ($B$ and $R$ filters) to acquire data starting at 16:47:41~UT, with a total exposure of 9.1~ks. During this observation, 1ES~1959+650 was detected with all three SVOM imaging instruments--ECLAIRs, MXT, and VT. The integrated 4--20~keV flux measured by ECLAIRs was approximately 2~mCrab, while MXT measured a 0.5--10~keV flux of $\sim$5.5~mCrab, well above the typical unabsorbed 0.3--10~keV flux reported by \citet{2016MNRAS.457..704K}. VT observations yielded optical magnitudes of $14.150 \pm 0.005$ and $13.787 \pm 0.004$ in the $VT_B$ and $VT_R$ filters, respectively (after correction for Galactic extinction using \citealt{1989ApJ...345..245C}). The MXT and VT measurements confirm that the source was in a significantly high state. 
   
   To monitor the evolution of the outburst, SVOM executed nine Target-of-Opportunity (ToO) observations between December 6 and December 26, 2024. Observations ceased after December 26 due to solar constraints that rendered the source inaccessible to the satellite \citep{2025ATel16978....1F}. \textit{Swift} initiated follow-up observations on December 12, 2024 \citep{2024ATel16941....1K, 2024ATel16955....1K}, six days after the initial SVOM detection, and continued monitoring until March 1, 2025, extending the temporal coverage of the flare. Table~\ref{tab:1ES_obs} summarizes all observations performed with SVOM and \textit{Swift}.

ntype Data from SVOM and \textit{Swift} were processed using their respective official pipelines. The SVOM/ECLAIRs spectra were generated by stacking all observations obtained on the same day to improve the signal-to-noise ratio (SNR) and were binned into 15 logarithmically spaced energy bins between 4 and 150 keV. The SVOM/MXT and \textit{Swift}/XRT spectra were rebinned to a minimum of 50 counts per bin using \texttt{ftgrouppha}, ensuring the validity of Gaussian statistics. The \textit{Swift}/UVOT and SVOM/VT magnitudes were extracted following the standard procedures for each instrument and converted into energy fluxes. Further details on the data reduction and analysis are provided in Appendix \ref{appendix:data}.

\section{Results} \label{sec:Results}
\subsection{Outburst evolution} \label{sec:outburst_evolution}

\begin{figure*}
        \centering
        \includegraphics[width=1.0\linewidth]{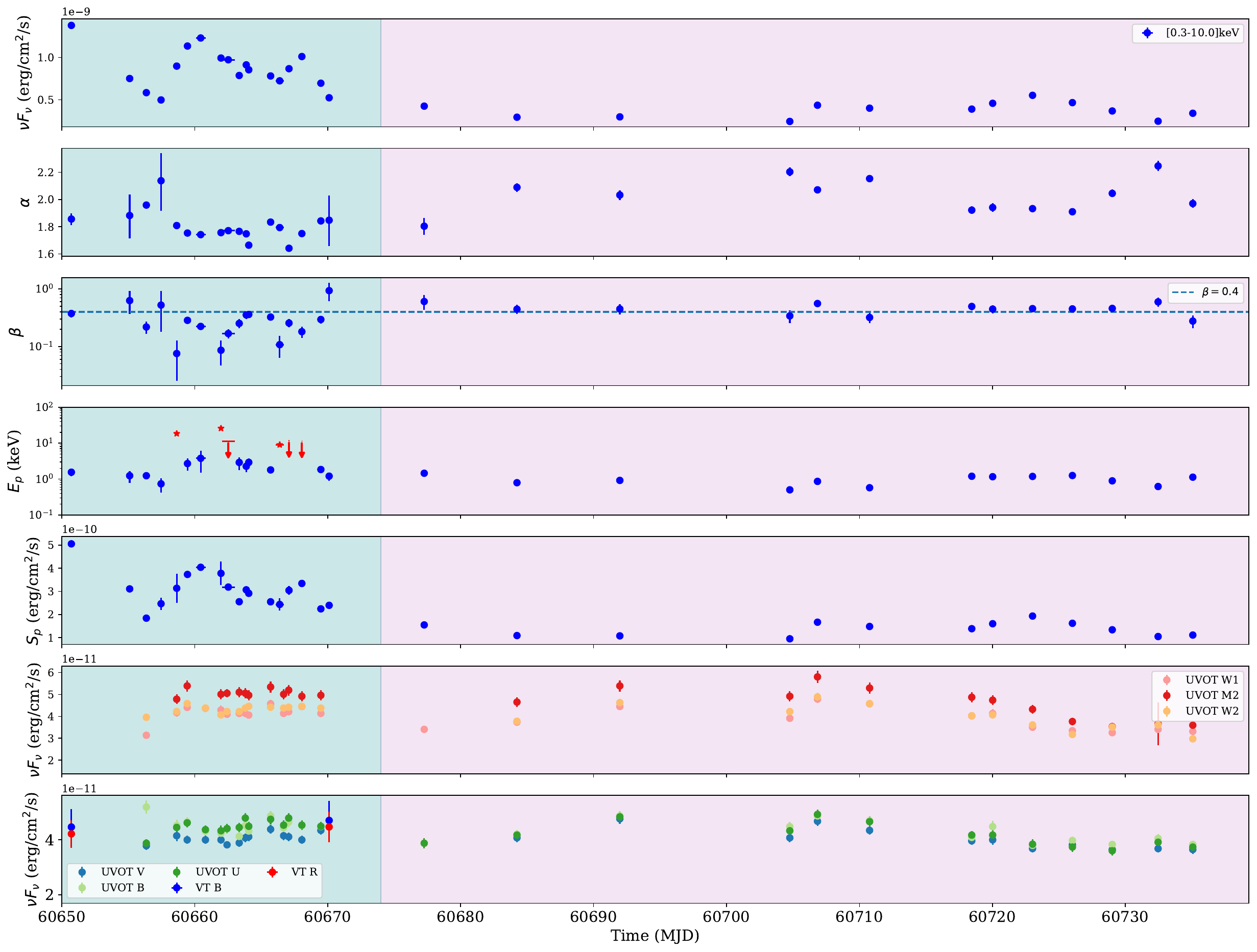}
        \caption{Evolution of the spectral parameters and the optical to X-ray fluxes. Are presented from top to bottom : intrinsic flux in the 0.3--10.0 keV band, the photon index, the curvature parameter, the energy of the SED synchrotron peak and the associated flux, the UV (W1, M2, W2 filters) and optical (V, B, U filters, as well as the VT R and B band) fluxes. The dashed blue line in the $\beta$ panel represent $\beta=0.4$, considered as a threshold value between low and high curvature. The red stars represent points that are unconstrained whereas red triangle represent $1\sigma$ upper limits. The blue and magenta backgrounds are respectively highlighting Period 1 and 2.}
        \label{fig:parameter_evol}
\end{figure*}

The evolution of the 0.3–10 keV energy flux as measured by SVOM/MXT, ECLAIRs, and \textit{Swift}/XRT, as well as the UV/optical flux measured by SVOM/VT and \textit{Swift}/UVOT, is shown in Fig.~\ref{fig:parameter_evol}.

For all days with simultaneous soft X-ray and ECLAIRs coverage, we performed joint spectral fits combining the MXT and/or XRT spectra with that of ECLAIRs. A multiplicative constant accounted for intercalibration offsets, using XRT as the reference (fixed to 1.0), while MXT and ECLAIRs constants were left free (see Table~\ref{tab:spectral_parameters}). After 2024 December 26, only XRT data were available due to SVOM operational constraints (see Table~\ref{tab:1ES_obs}), and only XRT spectra were fitted.

\begin{figure}
    \centering
    \includegraphics[width=0.9\linewidth]{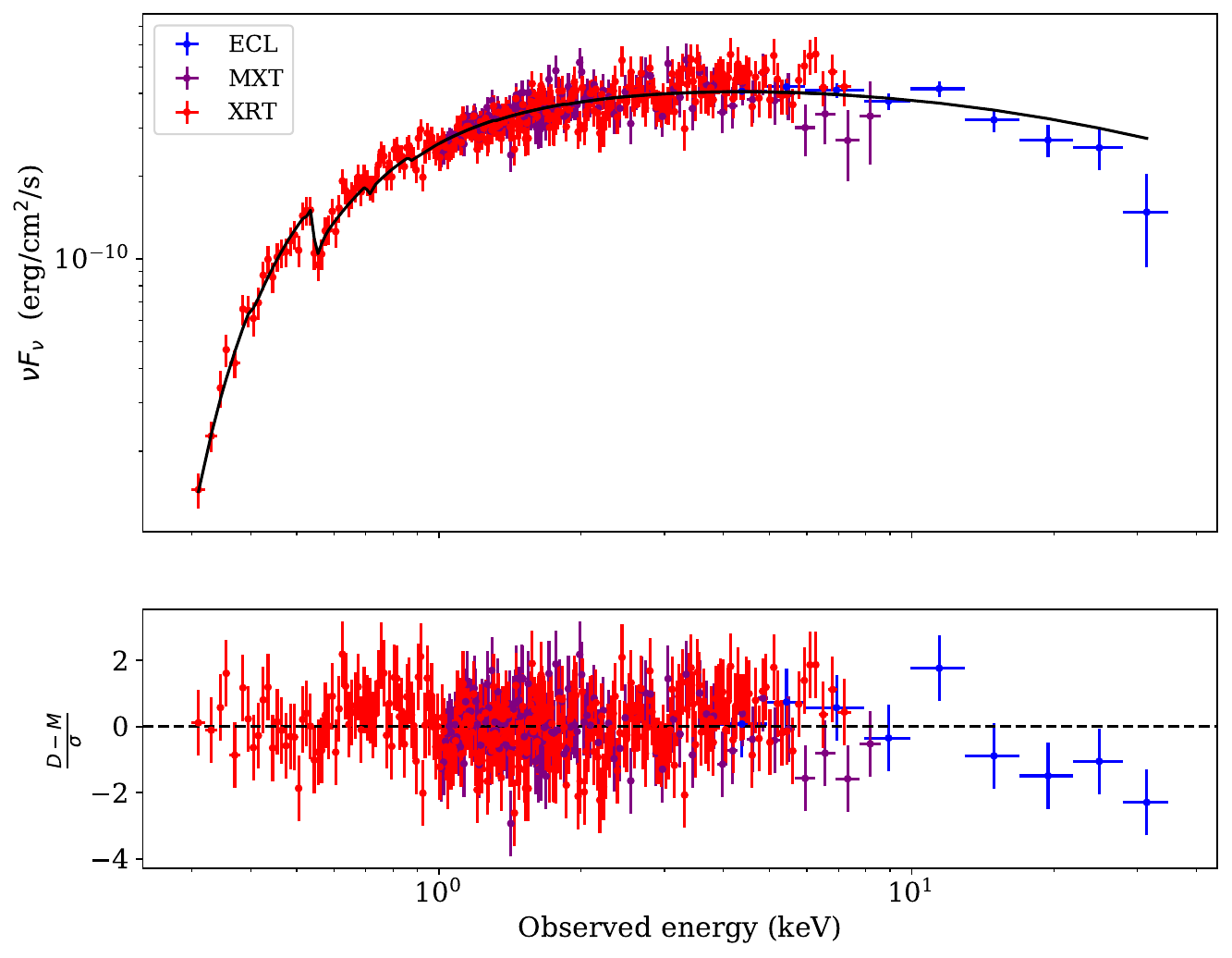}
    \caption{Example of joint-fit of the XRT, MXT and ECLAIRs for the observations taken on December 16, 2024. The bottom panel shows the residuals, computed as the difference between the observed data ($D$) and the model ($M$), normalized by the error ($\sigma$). }
    \label{fig:Spectra_ex}
\end{figure}

Spectra were modeled in \textsc{XSPEC} version~12.14.0 using an absorbed log-parabolic model (\texttt{constant*TBabs*logpar}; \citealt{2000ApJ...542..914W, 2004A&A...413..489M}), where \texttt{TBabs} accounts for Galactic absorption and the log-parabola is defined as
\begin{equation}
    F(E) = K \left( \frac{E}{E_1} \right)^{-(\alpha + \beta \cdot \log(E/E_1))} \quad 
    \mathrm{ph \, cm^{-2} \, s^{-1} \, keV^{-1}}.
\end{equation}
describes the intrinsic source spectrum. 
The pivot energy was fixed at $E_1 = 1$ keV and the absorption column density at the Galactic value $N_H = 1.0 \times 10^{21} \ \mathrm{cm^{-2}}$ \citep{2005A&A...440..775K}. The log-parabola model is described by three parameters: the photon index at 1 keV ($\alpha$), the curvature parameter $\beta$, and the normalization $K$. The fitted spectral parameters for all observations are listed in Table~\ref{tab:spectral_parameters}. An example of a joint fit for the 2024 December 16 observation is shown in Fig.~\ref{fig:Spectra_ex}.

From these parameters, we followed \citet{2004A&A...413..489M} and computed the synchrotron peak energy, $E_p$, and flux, $S_p$, as
\begin{align}
    E_p &= E_1 \cdot 10^{(2-\alpha)/(2\beta)} \ \mathrm{keV}, \\
    S_p &= (1.60 \times 10^{-9}) \ K \ E_1 \ 10^{(2-\alpha)^2/(4\beta)} \ \mathrm{erg \, cm^{-2} \, s^{-1}}.
\end{align}
Uncertainties (68\% confidence) on these two parameters were propagated from the fitted spectral parameters.

Fig.~\ref{fig:parameter_evol} shows the evolution of the X-ray and UV/optical energy flux of 1ES~1959+650 as observed throughout its outburst. The plot also shows the evolution of the fitted spectral parameters.

Two time periods can be distinguished from the 0.3–10~keV flux and $S_p$ levels. Period~1 (2024 December 6–26) shows higher fluxes, with weighted means $\bar{F}_{0.3-10 \ \mathrm{keV}} = (8.63 \pm 0.03) \times 10^{-10}$~erg~cm$^{-2}$~s$^{-1}$ and $\bar{S}_p = (2.50 \pm 0.01) \times 10^{-10}$~erg~cm$^{-2}$~s$^{-1}$. Period~2 (2025 January 2–March 1) shows lower values, $\bar{F}_{0.3-10 \ \mathrm{keV}} = (3.53 \pm 0.02) \times 10^{-10}$~erg~cm$^{-2}$~s$^{-1}$ and $\bar{S}_p = (1.35 \pm 0.01) \times 10^{-10}$~erg~cm$^{-2}$~s$^{-1}$.
The weighted mean photon index and curvature, $\bar\alpha$ and $\bar\beta$, are given in Table~\ref{tab:param_means}, with weights set by the relative uncertainties of the parameters.
    \begin{table}[]
        \caption{Weighted means values of the curvature parameter and the photon index during period 1 and 2 and during the total period.}
        \centering
        \begin{tabular}{l c c} 
            \hline
            \hline
            \textbf{Period}  & \textbf{$\bar\beta$} & \textbf{$\bar\alpha$} \\ 
            \hline
            Period 1 & $0.23\pm0.01$ & $1.77\pm0.01$ \\
            Period 2 & $0.44\pm0.02$ & $2.03 \pm 0.01$ \\
            Periods 1 and 2 & $0.27\pm0.01$ & $1.85\pm0.01$ \\ 
            \hline
        \end{tabular}
        \label{tab:param_means}
    \end{table}
    
\subsection{Spectral analysis}

To study the correlations between the parameters, we used the Spearman rank-order correlation coefficients, which are listed in Table~\ref{tab:param_correlations}. Period~1 is characterized by harder spectra with low curvature ($\bar\beta < 0.4$, the threshold distinguishing low- and high-curvature spectra) compared to Period~2 (see Table~\ref{tab:param_means}). The "harder when brighter" trend noted by \citet{2025ATel16978....1F} is confirmed by the strong anti-correlation between $\alpha$ and the 0.3--10~keV flux ($F_{0.3-10 \ \mathrm{keV}}$) across the full observation campaign (Table~\ref{tab:param_correlations}). Spectra also tend to exhibit lower curvature at higher fluxes, as suggested by an anti-correlation between $\beta$ and $F_{0.3-10 \ \mathrm{keV}}$. However, this trend is not statistically significant over the two periods.  

A shift of the synchrotron peak energy, $E_p$, toward higher energies is observed with increasing flux, yielding a positive correlation between $E_p$ and $F_{0.3-10 \ \mathrm{keV}}$ in both periods. Such correlations are often accompanied by an anti-correlation between X-ray and radio-to-UV flux, interpreted as a long-term change in the efficiency of the particle acceleration mechanism \citep{2006A&A...453...47K, 2018Galax...6..125K}. In our observations, no anti-correlation is seen between the X-ray flux and the optical flux in the V band ($F_V$), either within individual periods or in the combined dataset.  

The flare shows no significant $E_p-\beta$ anti-correlation (Table~\ref{tab:param_correlations}) in any period or in the combined dataset. Efficient stochastic acceleration is typically associated with low spectral curvature and a clear $E_p-\beta$ anti-correlation \citep{2011ApJ...739...66T, 2011ApJ...742L..32M}. In our case, only weak indications of this behavior are present, limited to low curvature during Period~1 and in the combined dataset (see Table~\ref{tab:param_means}). 

\begin{figure}
        \centering
        \includegraphics[width=0.9\linewidth]{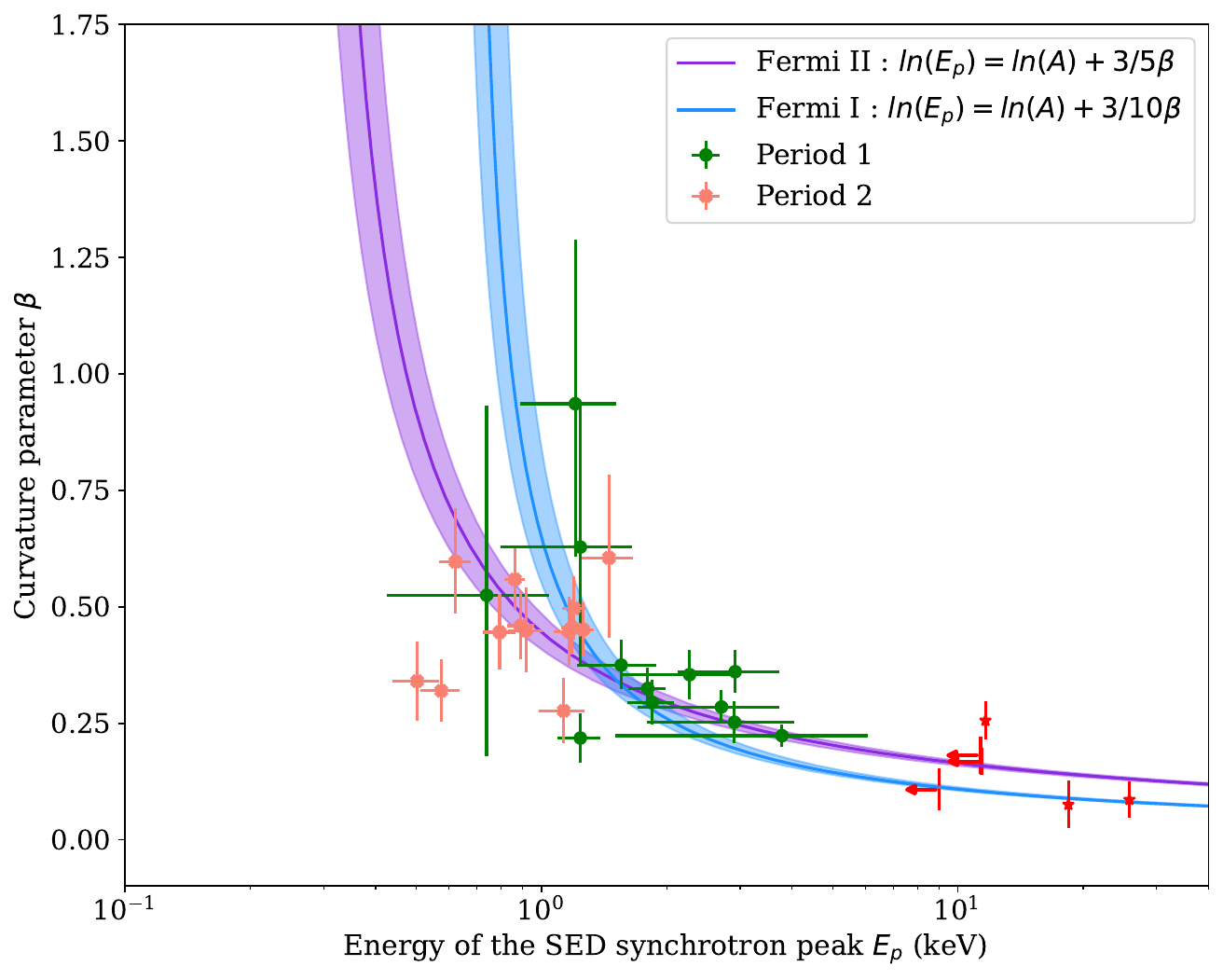}
        \caption{$E_p-\beta$ plan for the entire period (1+2). The purple and blue lines represent what we expect for Fermi II and Fermi I acceleration processes with an additional multiplicative constant fitted only using data from period 1. In both cases, the colored regions represent the error at the $1\sigma$ level. The red stars represent points that are unconstrained whereas red triangles represent $1\sigma$ upper limits. Errors on $E_{\rm p}$ at the $1\sigma$ level were obtained through error propagation using the $1\sigma$ uncertainties of the parameters $\alpha$ and $\beta$. When the lower bound of the confidence interval included zero, the value was treated as a $1\sigma$ upper limit. In addition, values were considered unconstrained when their uncertainties were at least one order of magnitude larger than the measured value itself, and when both limits of the confidence interval were poorly constrained.}
        \label{fig:Ep_b}
    \end{figure}

To further investigate the acceleration processes, we compared the data on the $E_p-\beta$ plane with theoretical predictions for Fermi I and Fermi II acceleration \citep{2011ApJ...739...66T, 2014ApJ...788..179C}, given by 
$\ln(E_p) = \ln(A) + 3/(10\beta)$ and $\ln(E_p) = \ln(A) + 3/(5\beta)$, respectively, with $A$ as a free parameter. Fits to Periods 1 and 2 combined yielded poor $\chi^2$ values, preventing firm conclusions regarding the dominant acceleration process. Fits on Period 1 alone were compatible with both Fermi I and Fermi II scenarios (see Fig.~\ref{fig:Ep_b}). For Period 1, the fits resulted in $\chi^2 = 17.9/17 = 1.05$ and $\chi^2 = 16.8/17 = 0.99$ for the Fermi I and Fermi II scenarios, respectively, while the combined fit to Periods 1 and 2 gave $\chi^2 = 104.1/30 = 3.47$ and $\chi^2 = 60.9/30 = 2.03$ for Fermi I and Fermi II, respectively. We also searched for a positive $\alpha-\beta$ correlation, expected for Fermi I \citep{2004A&A...413..489M}, but found none, either in individual periods or across the full outburst (Table~\ref{tab:param_correlations}).  

\begin{figure}
        \centering
        \includegraphics[width=0.9\linewidth]{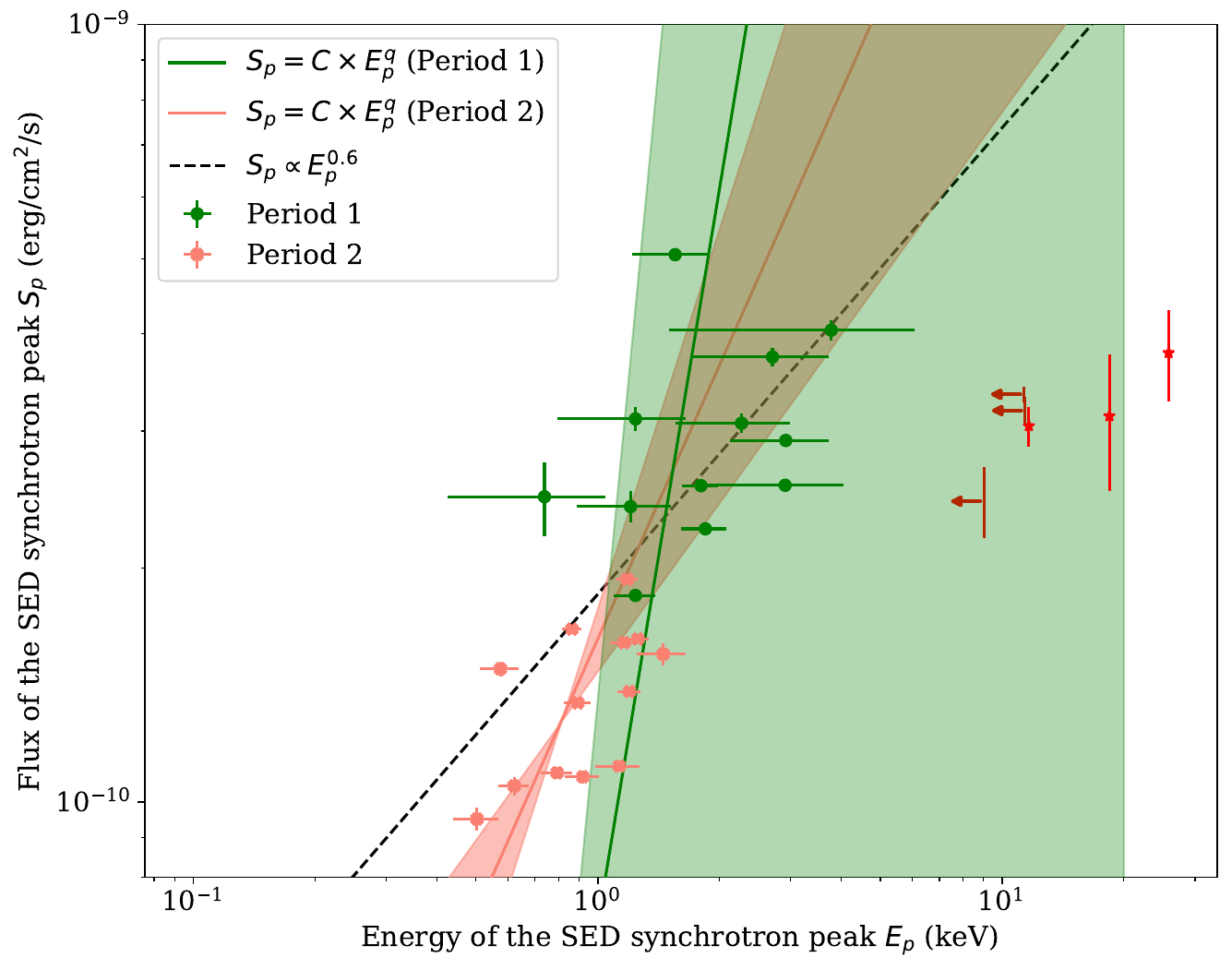}
        \caption{$S_p-E_p$ plan. The dashed line represents the case $S_p \propto E_p^{0.6}$. The orange and green lines represent the model $S_p = C \times E_p^q$ fitted to the data from period 1 and period 2, respectively. In both cases, the colored region represent the error at the $1\sigma$ level. The red stars represent points that are unconstrained whereas red triangles represent $1\sigma$ upper limits.}
        \label{fig:Sp_Ep}
    \end{figure}

A positive correlation between $E_p$ and $S_p$ was observed (Table~\ref{tab:param_correlations}), which is consistent with stochastic acceleration \citep{2011ApJ...739...66T}. Moreover, when the correlation follows $S_p \propto E_p^{0.6}$, it further indicates a transition in the turbulence energy spectrum, from a Kraichnan-type to a hard-sphere regime \citep{2011ApJ...739...66T, 2018Galax...6..125K}. We fitted a power-law model $S_p = C \, E_p^{q}$ (Fig.~\ref{fig:Sp_Ep}), finding $q = 3.14 \pm 2.27$, $q=1.17 \pm 0.45$, and $q=1.29 \pm 0.24$ respectively for Periods~1, 2, and both combined. Large uncertainties prevent a unique physical interpretation, as the results remain compatible with multiple scenarios \citep{2009A&A...501..879T}. 

\begin{figure}
        \centering
        \includegraphics[width=0.9\linewidth]{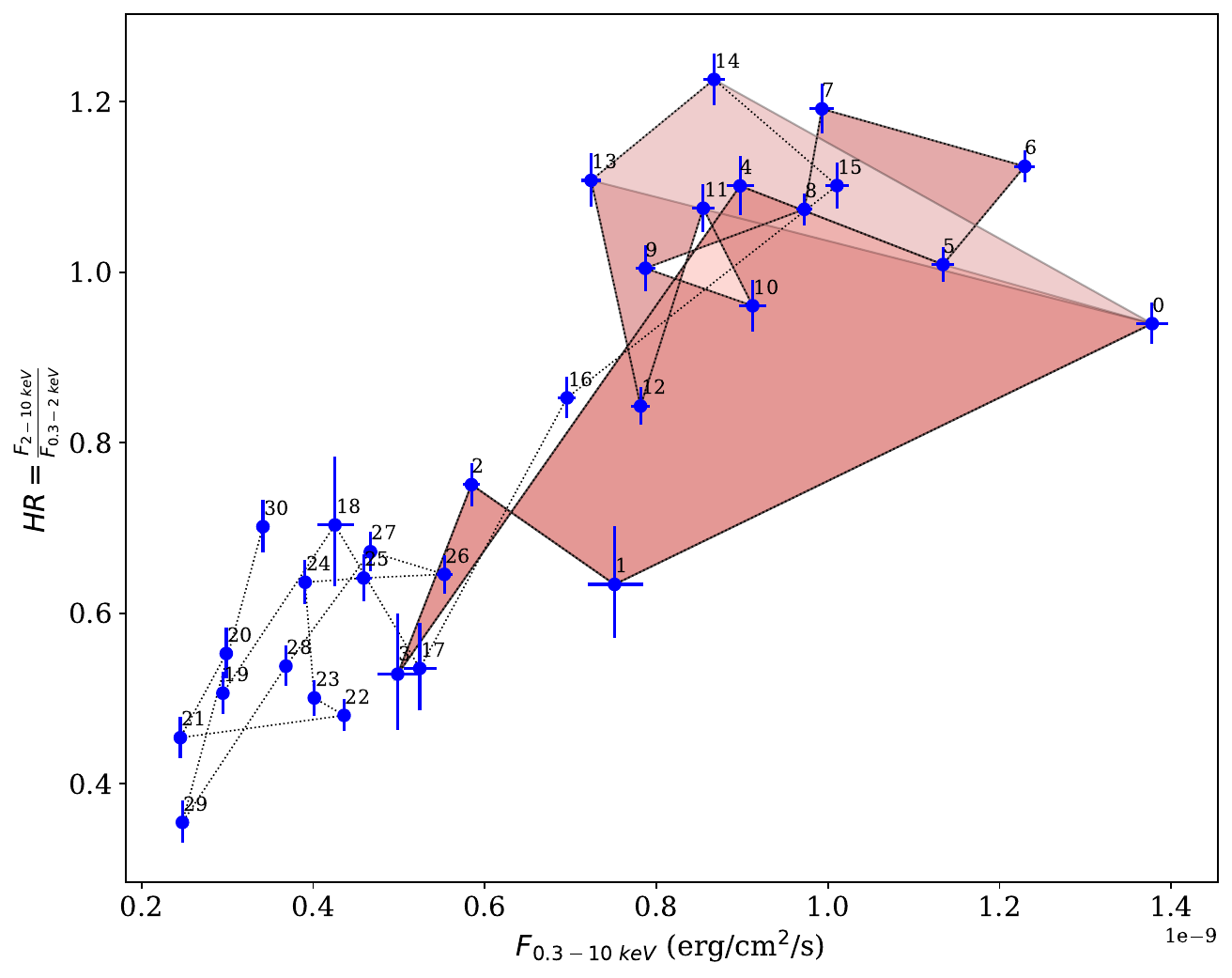}
        \caption{$HR-F_{0.3-10 \ \mathrm{keV}}$ plan for the entire follow-up campaign (period 1 and 2). The number near the data points represent the order to follow, ranging from 0 to 30. The areas shaded in orange represent the three loops initially identified and finally reject by the bootstrap algorithm. The three loops are : 0-6 CW, 0-14 CW; 0-15 CW. }
        \label{fig:HR_F}
\end{figure}

Finally, we examined the hardness ratio, $\mathrm{HR} = F_{2-10\,\mathrm{keV}} / F_{0.3-2\,\mathrm{keV}}$ \citep{2018ApJS..238...13K}, in the $\mathrm{HR}-F_{0.3-10\,\mathrm{keV}}$ plane (see Fig.~\ref{fig:HR_F}). In stochastic acceleration, gradual electron acceleration is expected to produce counter-clockwise (CCW) loops, whereas instantaneous Fermi~I acceleration may generate clockwise (CW) loops \citep[see e.g.][]{2004ApJ...605..662C,2005AIPC..801..410V}. A visual inspection of Fig.~\ref{fig:HR_F} reveals transitions between clockwise (CW) and counterclockwise (CCW) loops, a behavior commonly observed in HBLs and often interpreted as the result of shocks propagating through jet regions with varying physical conditions \citep{2018Galax...6..125K}. However, an assessment of the statistical significance of these loops (see Appendix~\ref{app:hysteresis} for details) indicates that none of them are statistically significant.

    \begin{table*}[]
        \caption{Spearman correlation coefficients computed for period 1 and period 2 as well as for the entire period.}
        \centering
        \begin{tabular}{l c c c c c c}
        \hline\hline
             \textbf{Parameters} & (1) \textbf{$\rho$ (Period 1 + Period 2)} & (2) \textbf{p-value} & (3) \textbf{$\rho$ (Period 1)} & (4) \textbf{p-value} & (5) \textbf{$\rho$ (Period 2)} & (6) \textbf{p-value} \\
             \hline
             
             $\alpha-F_{0.3-10 \ \mathrm{keV}}$ & $-0.83$ &  $\mathbf{1.04\times10^{-8}}$ & $-0.54$ & $0.02$ & $-0.77$ & $\mathbf{0.002}$ \\
             
             $\beta-F_{0.3-10 \ \mathrm{keV}}$ & $-0.52$ & $0.03$ & $-0.28$ & $0.26$ & $0.15$ & $0.45$ \\
             $E_p-\beta$ & $-0.38$ & $0.06$ & $-0.53$ & $0.07$ & $0.21$ & $0.20$ \\
             
             $F_{V}-F_{0.3-10 \ \mathrm{keV}}$ & $0.08$ & $0.71$ & $-0.32$ & $0.26$ & $-0.01$ & $0.97$ \\
             $S_p-E_p$ & $0.78$ &  $\mathbf{4.59\times10^{-06}}$ & $0.48$ & $0.11$ & $0.58$ & $0.04$ \\
             
             $E_p-F_{0.3-10 \ \mathrm{keV}}$ & $0.84$ &  $\mathbf{1.67\times10^{-07}}$ & $0.64$ & $0.02$ & $0.72$ & $0.004$ \\
             
             $\alpha-\beta$ & $0.49$ & $0.005$ & $0.31$ & $0.21$ & $-0.08$ & $0.36$ \\
             \hline
        \end{tabular}
        \tablefoot{Spearman correlation coefficients computed for period 1 and period 2 as well as for the entire period. \textit{Column 1-2 :} Spearman correlation coefficient for Periods 1 and 2 with the associated p-values. \textit{Column 3-4 :} Spearman correlation coefficient for Periods 1 with the associated p-values. \textit{Column 5-6 :} Spearman correlation coefficient for Periods 2 with the associated p-values. The correlation coefficients between $E_p$ and another parameter do not take into account the upper limits or the unconstrained values. P-values in bold indicate correlations with a significance greater than $3\sigma$.}
        \label{tab:param_correlations}
    \end{table*}

\section{Discussions and conclusions} 

On December 6, 2024, the ECLAIRs telescope on board SVOM detected a long X-ray transient spatially coincident with the blazar 1ES~1959+650, marking the first blazar outburst observed by SVOM \citep{2024ATel16935....1C}. Following the trigger, SVOM performed a ToO monitoring campaign until December 26, 2024, capturing the spectral evolution of the source. \textit{Swift} complemented these observations with 27 pointings from six days after the first detection until 2025 March 1 \citep{2024ATel16941....1K, 2024ATel16955....1K}. The spectral evolution of the source was modeled using a log-parabola to identify possible correlations and tentatively probe the flare’s acceleration mechanisms.

The flare exhibits a clear harder-when-brighter trend, along with a shift of the synchrotron peak energy ($E_p$) toward higher energies, as indicated by the anti-correlation between $\alpha$ and $F_{0.3-10\,\mathrm{keV}}$ and the positive correlation $E_p-F_{0.3-10\,\mathrm{keV}}$. No corresponding decrease in the optical-to-UV flux was observed during the X-ray rise. Neither Fermi II nor Fermi I signatures were clearly detected: fitting the data to theoretical models in the $E_p-\beta$ plane were inconclusive, and no significant $\alpha-\beta$ correlation was observed. A positive $S_p-E_p$ correlation was measured, consistent with stochastic acceleration, but the exponent remained poorly constrained. Similarly, no significant loops were identified in the $\mathrm{HR}-F_{0.3-10\,\mathrm{keV}}$ plane, providing no clear indication of the dominant acceleration process (see Appendix~\ref{app:hysteresis}).

The absence of strong signatures from either Fermi I or Fermi II acceleration suggests that both processes may operate concurrently. Particles could undergo initial shock acceleration followed by stochastic re-acceleration downstream, or experience multiple shock crossings \citep{2006A&A...453...47K, 2012SSRv..173..535P}. Consequently, the combination of these mechanisms may weaken individual signatures. Compared to previous flares of 1ES~1959+650, this event shows no clear Fermi-I or Fermi-II dominance, resembling the 2015–2016 outburst \citep{2016MNRAS.461L..26K}, whereas other flares have shown clearer stochastic acceleration signatures \citep[see e.g.][]{2018ApJS..238...13K}.

This study highlights the first blazar outburst detection by SVOM, demonstrating the rapid response of the satellite and collaboration with follow-up observations starting less than two hours after the trigger. The multi-day SVOM campaign, jointly with \textit{Swift}, illustrates the SVOM mission’s potential for continuous monitoring of blazar activity, providing dense temporal and spectral coverage. In particular, the low-energy threshold of ECLAIRs ensures spectral continuity beyond the 10~keV limit covered by XRT and MXT. This yields tighter constraints on physical parameters. Future SVOM observations of 1ES~1959+650, together with multi-wavelength facilities will enhance temporal and spectral sampling, offering new opportunities to probe particle acceleration mechanisms in blazar flares.

\begin{acknowledgements}
      This work was supported by CNES, focused on SVOM. The Space-based multi-band Variable Objects Monitor (SVOM) is a joint Chinese-French mission led by the Chinese National Space Administration (CNSA), the French Space Agency (CNES), and the Chinese Academy of Sciences (CAS). We gratefully acknowledge the unwavering support of NSSC, IAMCAS, XIOPM, NAOC, IHEP, CNES, CEA, CNRS, University of Leicester, and MPE.We would like to thank the Swift team for carrying out the observations we proposed. In addition to our own data, we have also used the Swift archive at \url{https://swift.gsfc.nasa.gov/archive/}. SK would like to thank the CAS President's International Fellowship Initiative for visiting scientists.
\end{acknowledgements}

\bibliographystyle{aa} 
\bibliography{biblio.bib}

\begin{appendix} 

\section{Instruments and data reduction}\label{appendix:data}

    \subsection{The SVOM space mission}
        SVOM has discovered 1ES 1959+650 on December 6, 2024, in an outburst phase with the ECLAIRs telescope \citep{2024ATel16935....1C} onboard the satellite. The outburst phase of 1ES 1959+650 has then been followed through the Target of Opportunity (ToO) program until December 26, 2024 (\cite{2025ATel16978....1F}, see Table~\ref{tab:1ES_obs}). Because of solar constraints, SVOM has not been able to observe the entire flaring phase. Note that the observations have been performed during the commissioning and verification phase of SVOM.
    
    \subsubsection{The ECLAIRs (4--150\,keV) instrument}
        The ECLAIRs telescope (\citealt{2014SPIE.9144E..24G}, Godet et al. in prep) is a coded mask instrument with a large field of view (FoV) of $\sim2$ sr, operating in the 4–150 keV energy band. The data have been reduced using the official pipeline, namely ECPI version (Goldwurm et al. in prep), with the calibration files from version 20240101. 
        
        The ECPI pipeline reduces the ECLAIRs data based on the following steps. Good Time Intervals (GTIs) are defined, and energy calibration of the events is performed by applying a pixel gain and offset, while events that are not valid are flagged. Binned detector images are then created and corrected for efficiency, non-uniformity, and background. Sky images are finally reconstructed through deconvolution and coding-noise cleaning. Based on the source position in the sky images, ECPI builds a specific shadowgram model, which is fitted to the detector maps to extract the flux in each energy bin, while a second-degree polynomial accounts for the background component.
        
        We considered all the observations in which 1ES 1959+650 was detected. For each observation considered, a spectrum was produced with 15 bins logarithmically distributed over the entire ECLAIRs energy band. The response matrix file and the ancillary response file used were respectively ECL-RSP-RMF\_20220515T01.fits and ECL-RSP-ARF\_20220515T01.fits. To increase the signal-to-noise ratio of the spectrum, we then combined the spectra from observations taken on the same day.

    \subsubsection{The MXT (0.3--10\,keV) instrument}
        The MXT telescope (Microchannel X-ray Telescope, \citealt{2023ExA....55..487G}, Götz et al. in prep) has a $58\times58$ arcmin squared FoV, operating in the 0.3–10 keV energy band. For every observation, the data have been reduced and the spectra produced using the official pipeline, called MXT-pipeline (Maggi et al. in prep), in its 1.9.0 version. The response matrix file and ancillary response file used in the analysis were MXT-GND-MATRIX-ALL\_20250206T093900\_20240315113300.2.fits and MXT\_FM\_PANTER\_FULL-ALL-1.1.arf. The background file used was produced with the MXT-pipeline task. Because of instrumental calibration issues, we chose to consider the data only above 1.0 keV. Also, to ensure the possibility of using the Gaussian approximation, we binned the spectra with at least 50 cts/bin using \texttt{ftgrouppha}, developed as part of the \texttt{HEASoft} suite.

    \subsubsection{The VT optical telescope}
        The VT telescope (Visible Telescope, \citealt{2020SPIE11443E..0QF}, Qiu et al. in prep) is an optical telescope (40~cm aperture size) operating with two filters: the $VT_B$ (400–650 nm) and $VT_R$ (650–1000 nm) bands, with a field of view (FoV) of 26' in diameter. The data were analyzed following the standard procedure. 1ES 1959+650 has been observed by the VT only for two observations, taken on December 6 and December 26, 2024 \citep{2025ATel16978....1F}.

        The magnitudes were measured in the AB reference system. They were corrected for extinction following \citet{1989ApJ...345..245C}. The corrected magnitudes were then converted into energy fluxes by adopting effective wavelengths of $\lambda_R = 789.6 \pm 94.0$~nm and $\lambda_B = 533.7 \pm 78.5$~nm for the R and B filters, respectively. Flux uncertainties were computed via error propagation, taking into account the uncertainties on the magnitudes as well as on the reference wavelengths for each filter. An error of 5\% has been assumed on the magnitude values (internal communication). 

    \subsection{SVOM trigger and alert system}

        On Friday, December 6 at 15:08:48~UT ($T_b$), the on-board trigger software of the ECLAIRs telescope detected and localized a long-duration soft X-ray transient at $\mathrm{RA} = 300.202^\circ$, $\mathrm{DEC} = 65.182^\circ$ (J2000). The source position is constrained within a $9.3$~arcmin uncertainty radius in the 5--8~keV energy band, during a 22-min exposure starting at $T_b$. A subsequent alert with SNR = 7.2 was issued in the same band during an 11-minute exposure beginning at 15:14:15~UT. The sub-image transmitted in near real-time via the SVOM VHF network revealed a clear point-like source absent from the on-board catalog.  
        
        Following this trigger, a Target of Opportunity (ToO) observation with MXT and VT was carried out at 16:47:41~UT, with an exposure of 3.6~ks. A single source was detected at $\mathrm{RA} = 300.0125^\circ$, $\mathrm{DEC} = 65.1519^\circ$ (J2000), with a positional uncertainty of $25$~arcsec (90\% confidence level). This source lies $22$~arcsec from the BL Lac blazar 1ES~1959+650.

    \subsection{The Swift Neil Gehrels Observatory}
 
        1ES 1959+650 was followed up with the Neil Gehrels Swift Observatory (\citealt{2004ApJ...611.1005G}) in the period December 12, 2024 to March 01, 2025, starting 6 days after the initial SVOM discovery of the X-ray outburst (\citealt{2024ATel16941....1K, 2024ATel16955....1K}, see Table~\ref{tab:1ES_obs}). In addition to our own \textit{Swift} follow-ups, we also analyzed archival \textit{Swift} observations from that epoch. 
        For comparison with the ongoing outburst, we also analyzed earlier \textit{Swift} observations from August 2024 when the blazar was found to be in its intermediate-to-low state. 
    
        \subsubsection{The XRT (0.3--10\,keV) telescope}

        The \textit{Swift} X-ray telescope (XRT) observations were carried out in windowed timing (WT) mode because of the high X-ray count rate. Data were reduced with the \textsc{xrtpipeline} 0.13.7, that is part of the \texttt{HEASoft} package 6.35.1. For spectral analysis the response file swxwt0s6\_20210101v016.rmf was used. Source counts were extracted in a rectangular region of radius 40x3 pixels. Source spectra of each epoch were produced and fitted with a single powerlaw model with absorption fixed at the Galactic value of $1.0\times10^{21} \ \mathrm{cm^{-2}}$ (\cite{2005A&A...440..775K}), and free photon index $\Gamma$.  
        
\begin{figure}
    \centering
    \includegraphics[width=1.0\linewidth]{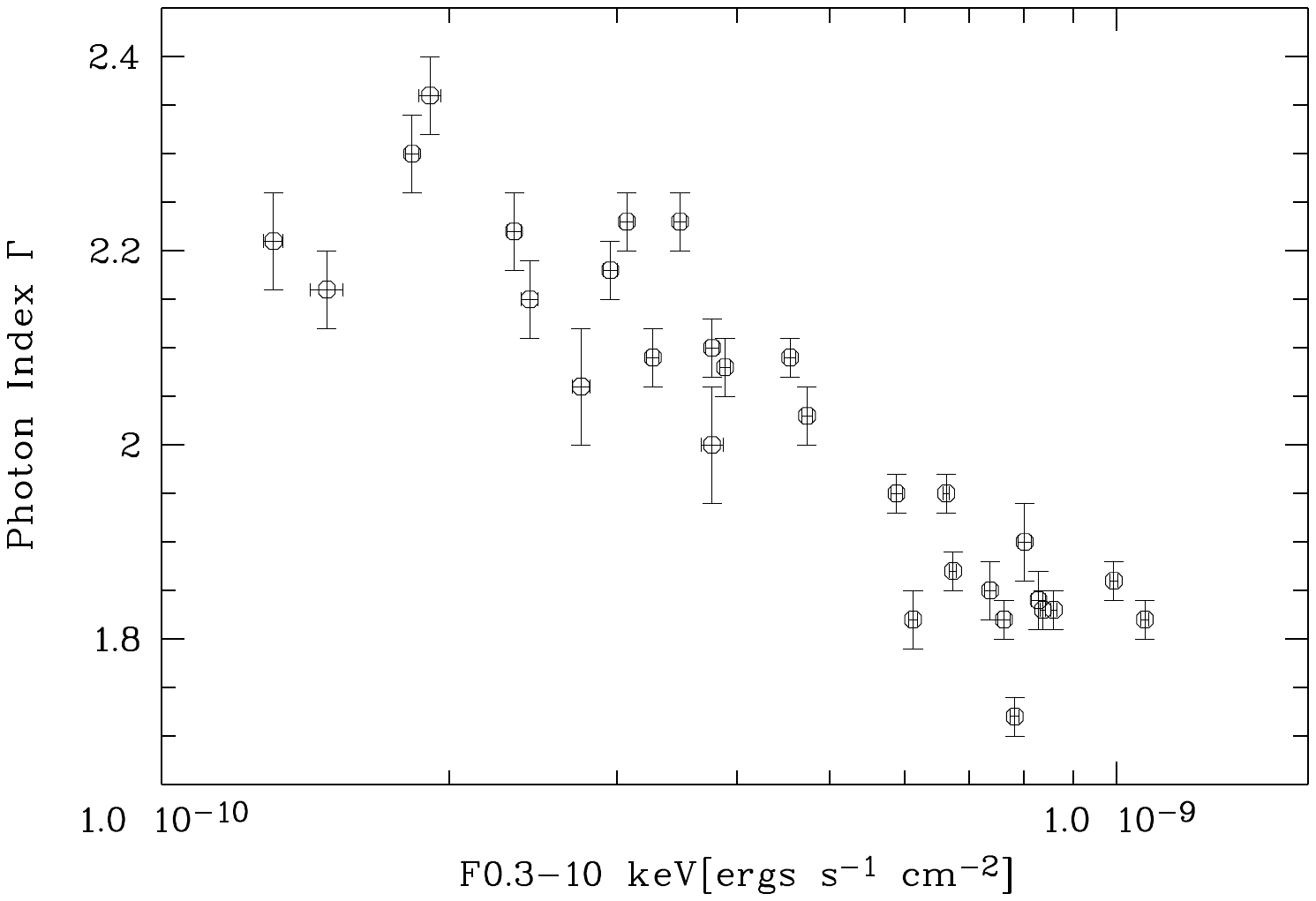}
    \caption{Power-law photon index $\Gamma$ as function of the 0.3-10 keV flux of the XRT observations of 1ES 1959+650 between August 2024 and March 2025. The Spearman correlation coefficient $\rho=-0.89$ determined with a p-value $p=5\times10^{-11}$.}
    \label{fig:Gamma_flux_xrt}
\end{figure}

        During the Swift observations, the X-ray count rate varied by a factor $\sim4.6$ and the peak flux was reached on December 16, 2024. A very strong anti-correlation between X-ray flux and spectral index $\Gamma$ is immediately evident (see Fig.~\ref{fig:Gamma_flux_xrt}). For further analysis, the \textit{Swift}/XRT and SVOM X-ray data were combined. For the following analysis, to ensure the possibility to use a Gaussian approximation, we binned the spectra with at least 50 cts/bin using \texttt{ftgrouppha}. 

        \subsubsection{The UVOT (UV/optical) telescope}

            1ES 1959+650 was also observed with the \textit{Swift} UV-optical telescope (UVOT) in all three optical and three UV filters in order to ensure good SED coverage of this highly variable blazar. The UVOT data of each segment were co-added using the task \textsc{uvotimsum}. Source counts were extracted in a circular region of radius 5~arcseconds. UVOT magnitudes were measured following standard procedures and were corrected for Galactic extinction based on \cite{1989ApJ...345..245C}. The optical--UV emission was also found to be in a high state. However, it did not closely follow the X-ray evolution. The optical maximum was reached on January 31, 2025, which is 56 days after the observed X-ray peak on December 6, 2024.

\begin{table*}[]
    \caption{Summary of the joint-observation campaign of SVOM and Swift for the outburst phase of 1ES 1959+650 from December 6, 2024 to March 01, 2025.}
    \centering
    \begin{tabular}{l c c c c c}
    \hline
    \hline
        \textbf{Date} (1) & \textbf{SVOM ObsID} (2) & \textbf{MXT $t_{\rm{exp}}$ (s)} (3) & \textbf{XRT ObsID} (4) & \textbf{XRT $t_{\rm{exp}}$ (s)} (5) & \textbf{ECLAIRs $t_{\rm{exp}}$ (s)} (6) \\ \hline
        24/12/06 & 1426066399 & 9157.4 & --- & --- & 16635.2 \\
        24/12/11 & 1140853734 & 2084.5 &  --- & --- & 6240 \\
        24/12/12 & --- & --- & 00013906125 & 903.1 & --- \\
        24/12/13 & 1140853773 & 2083.0 & --- & --- & 1365 \\
        24/12/14 & --- & --- & 00013906126 & 674.1 & 1449 \\
        24/12/15 & 1140853775 & 2079.3 & 00013906129 & 942.0 & 1411.1\\
        24/12/16 & 1140853776 & 2085.5 & 00013906131 & 838.8 & 5695 \\
        24/12/17 & --- & --- & 00013906132 &  917.3 & --- \\
        24/12/18 & 1140853785 & 3786.2 & 00013906133 &  1820.8 & 6727.1 \\
        24/12/19 AM & --- & --- & 00013906134 &  908.1 & --- \\
        24/12/19 PM & 1140853847 & 3535.1 & 00013906135 & 908.1 & 9848.1 \\
        24/12/20 & --- & --- & 00013906136 & 1022.5 & --- \\
        24/12/21 & --- & --- & 00013906137 & 938.2 & --- \\
        24/12/22 & 1140853853 & 1529.1 & 00013906138 & 918.1 & 3059.1 \\
        24/12/23 & 1140853852 & 1436.0  & 00013906139 &  855.6 & 5349.3\\
        24/12/24 & --- & --- & 00013906140 & 928.1 & --- \\
        24/12/25 & --- & --- & 00013906141 & 929.2 & --- \\
        24/12/26 & 1140853913 & 3016.75 & --- & --- & 3300.1 \\
        25/01/02 & --- & --- & 00013906142 & 209.4 & --- \\
        25/01/09 & --- & --- & 00013906144 & 863.2 & --- \\
        25/01/16 & --- & --- & 00013906145 & 703.3 & --- \\
        25/01/29 & --- & --- & 00013906147 & 1117.8 & --- \\
        25/01/31 & --- & --- & 00013906148 & 917.8 & --- \\
        25/02/04 & --- & --- & 00013906149 & 893.1 & --- \\
        25/02/12 & --- & --- & 00013906151 & 908.1 & --- \\
        25/02/14 & --- & --- & 00013906152 & 666.3 & --- \\
        25/02/17 & --- & --- & 00013906153 & 924.3 & --- \\
        25/02/20 & --- & --- & 00013906154 & 952.2 & --- \\
        25/02/23 & --- & --- & 00013906155 & 883.1 & --- \\
        25/02/26 & --- & --- & 00013906156 & 738.6 & --- \\
        25/03/01 & --- & --- & 00013906157 & 839.4 & --- \\
        \hline
    \end{tabular}
     \tablefoot{\textit{Column (1)} represent the day where the observations were performed. \textit{Column (2-3)} give the SVOM ObsID and MXT exposure time (in seconds) respectively. \textit{Column (4-5)} give the XRT ObsID and exposure time (in seconds). \textit{Column (6)} represent the exposure of the stacked ECLAIRs spectrum of the day. Note that on December 19, 2024, two XRT observations were performed, that is why we distinguish them for this day.}
    \label{tab:1ES_obs}
\end{table*}

\section{Long-term spectral parameter evolution}

Table~\ref{tab:spectral_parameters} provides the fitted spectral parameters for each dataset considered in our work. 
As explained in Section~\ref{sec:outburst_evolution}, the outburst phase can be divided into two distinct periods. The first period (2024 December 6–26), covered by both SVOM and \textit{Swift}, shows a high flux level in the 0.3–10~keV band ($\bar{F}_{0.3-10 \ \mathrm{keV}} = (8.63 \pm 0.03) \times 10^{-10}$~erg~cm$^{-2}$~s$^{-1}$), with a hard spectrum combined with low curvature ($\bar{\beta}<0.4$, see Table~\ref{tab:param_means}). In contrast, the second period (2025 January 2–March 1), covered solely by \textit{Swift}, exhibits a lower spectral flux in the same energy band ($\bar{F}_{0.3-10 \ \mathrm{keV}} = (3.53 \pm 0.02) \times 10^{-10}$~erg~cm$^{-2}$~s$^{-1}$), along with a softer spectrum and higher curvature ($\bar{\beta}>0.4$). Meanwhile, the optical and UV fluxes, computed in the V and W1 bands respectively, show slightly higher values during period~2 ($\bar F_V = (4.07\pm0.15)\times10^{-11}$~erg~cm$^{-2}$~s$^{-1}$ and $\bar F_{W1} = (3.83\pm0.05)\times10^{-11}$~erg~cm$^{-2}$~s$^{-1}$) compared to period~1 ($\bar F_V = (3.59\pm0.03)\times10^{-11}$~erg~cm$^{-2}$~s$^{-1}$ and $\bar F_{W1} = (3.27\pm0.03)\times10^{-11}$~erg~cm$^{-2}$~s$^{-1}$). 


 \begin{table*}[]
    \caption{Spectral parameters.}
     \centering
     \resizebox{\textwidth}{!}{
         \begin{tabular}{l c c c c c c c c}
         \hline
         \hline
            \textbf{Date} (1) &  \textbf{$\alpha$} (2) & \textbf{$\beta$} (3) & \textbf{$F_{0.3-10\,\mathrm{keV}}$} (4) & \textbf{$E_p$ [keV]} (5) & \textbf{$S_p$} (6) & \textbf{$C_{MXT}$} (7) & \textbf{$C_{ECL}$} (8) & \textbf{$\chi^2_r/\mathrm{dof}$} (9) \\
            \hline
            24/12/06 & $1.856^{+0.043}_{-0.044}$ & $0.375^{+0.054}_{-0.052}$ & $1.378^{+0.019}_{-0.018}$ & $1.55^{+0.34}_{-0.34}$ & $5.06^{+0.10}_{-0.10}$ & --- & $1.15^{+0.07}_{-0.06}$ & $1.20/197$ \\
            
            24/12/11 & $1.883^{+0.168}_{-0.154}$ & $0.628^{+0.301}_{-0.256}$ & $0.751^{+0.034}_{-0.031}$ & $1.24^{+0.41}_{-0.44}$ & $3.11^{+0.11}_{-0.11}$ & --- & $1.15^{+0.29}_{-0.23}$ & $0.80/72$ \\
            
            24/12/12 & $1.959^{+0.023}_{-0.023}$ & $0.218^{+0.053}_{-0.054}$ & $0.584^{+0.010}_{-0.010}$ & $1.24^{+0.14}_{-0.15}$ & $1.84^{+0.02}_{-0.02}$ & --- & --- & $1.05/163$ \\
            
            24/12/13 & $2.138^{+0.201}_{-0.220}$ & $0.525^{+0.406}_{-0.346}$ & $0.498^{+0.025}_{-0.023}$ & $0.74^{+0.30}_{-0.31}$ & $2.47^{+0.27}_{-0.27}$ & --- & $0.89^{+0.68}_{-0.56}$ & $1.13/60$ \\
            
            24/12/14 & $1.809^{+0.025}_{-0.025}$ & $0.076^{+0.050}_{-0.051}$ & $0.897^{+0.016}_{-0.016}$ & $18.42^ \dagger$ & $3.14^{+0.63}_{-0.62}$ & --- & $0.96^{+0.15}_{-0.14}$ & $1.05/170$ \\
            
            24/12/15 & $1.754^{+0.018}_{-0.018}$ & $0.285^{+0.036}_{-0.035}$ & $1.134^{+0.013}_{-0.013}$ & $2.70^{+1.01}_{-1.01}$ & $3.73^{+0.10}_{-0.10}$ & $1.25^{+0.02}_{-0.02}$ & $1.08^{+0.14}_{-0.14}$ & $1.07/353$ \\
            
            24/12/16 & $1.742^{+0.017}_{-0.017}$ & $0.223^{+0.024}_{-0.024}$ & $1.229^{+0.012}_{-0.012}$ & $3.78^{+2.30}_{-2.28}$ & $4.04^{+0.12}_{-0.12}$ & $1.90^{+0.03}_{-0.03}$ & $2.00^{+0.09}_{-0.08}$ & $0.94/380$ \\
            
            24/12/17 & $1.757^{+0.020}_{-0.020}$ & $0.086^{+0.040}_{-0.040}$ & $0.993^{+0.014}_{-0.014}$ & $25.81^\dagger$ & $3.78^{+0.51}_{-0.50}$ & --- & --- & $1.07/227$ \\
            
            24/12/18 & $1.771^{+0.015}_{-0.016}$ & $0.168^{+0.029}_{-0.029}$ & $0.972^{+0.010}_{-0.010}$ & $<11.4$ & $3.19^{+0.14}_{-0.14}$ & $0.82^{+0.02}_{-0.01}$ & $0.60^{+0.07}_{-0.07}$ & $1.14/398$ \\
            
            24/12/19 AM & $1.766^{+0.022}_{-0.022}$ & $0.252^{+0.045}_{-0.045}$ & $0.787^{+0.012}_{-0.012}$ & $2.91^{+1.14}_{-1.12}$ & $2.55^{+0.03}_{-0.04}$ & --- & --- & $1.04/205$ \\
            
            24/12/19 PM & $1.750^{+0.027}_{-0.027}$ & $0.354^{+0.054}_{-0.052}$ & $0.912^{+0.016}_{-0.015}$ & $2.27^{+0.72}_{-0.71}$ & $3.07^{+0.09}_{-0.09}$ & $1.01^{+0.02}_{-0.02}$ & $0.93^{+0.10}_{-0.09}$ & $1.13/255$ \\
            
            24/12/20 & $1.664^{+0.023}_{-0.023}$ & $0.361^{+0.046}_{-0.047}$ & $0.855^{+0.013}_{-0.013}$ & $2.92^{+0.81}_{-0.79}$ & $2.92^{+0.04}_{-0.04}$ & --- & --- & $0.88/209$ \\
            
            24/12/21 & $1.834^{+0.021}_{-0.021}$ & $0.325^{+0.044}_{-0.045}$ & $0.782^{+0.011}_{-0.011}$ & $1.81^{+0.19}_{-0.19}$ & $2.55^{+0.03}_{-0.03}$ & --- & --- & $0.98/211$ \\
            
            24/12/22 & $1.794^{+0.023}_{-0.023}$ & $0.108^{+0.045}_{-0.044}$ & $0.724^{+0.012}_{-0.012}$ & $9.03^\dagger$ & $2.44^{+0.21}_{-0.21}$ & $1.54^{+0.04}_{-0.04}$ & $1.15^{+0.13}_{-0.13}$ & $1.19/235$ \\
            
            24/12/23 & $1.642^{+0.022}_{-0.022}$ & $0.256^{+0.042}_{-0.041}$ & $0.867^{+0.013}_{-0.013}$ & $<11.45$ & $3.05^{+0.18}_{-0.18}$ & $1.06^{+0.03}_{-0.03}$ & $0.84^{+0.09}_{-0.09}$ & $1.27/269$ \\
            
            24/12/24 & $1.750^{+0.020}_{-0.020}$ & $0.181^{+0.040}_{-0.040}$ & $1.010^{+0.014}_{-0.014}$ & $<11.34$ & $3.34^{+0.08}_{-0.08}$ & --- & --- & $1.19/232$ \\
            
            24/12/25 & $1.843^{+0.022}_{-0.022}$ & $0.295^{+0.047}_{-0.048}$ & $0.696^{+0.010}_{-0.010}$ & $1.85^{+0.24}_{-0.24}$ & $2.25^{+0.03}_{-0.03}$ & --- & --- & $1.09/193$ \\
            
            24/12/26 & $1.848^{+0.183}_{-0.188}$ & $0.936^{+0.328}_{-0.351}$ & $0.524^{+0.020}_{-0.019}$ & $1.36^{+0.31}_{-0.31}$ & $2.40^{+0.11}_{-0.11}$ & --- & $1.20^{+0.48}_{-0.41}$ & $0.91/75$ \\
            
            25/01/02 & $1.804^{+0.065}_{-0.062}$ & $0.605^{+0.172}_{-0.179}$ & $0.425^{+0.022}_{-0.020}$ & $1.45^{+0.20}_{-0.20}$ & $1.55^{+0.05}_{-0.05}$ & --- & --- & $1.12/33$ \\
            
            25/01/09 & $2.089^{+0.032}_{-0.032}$ & $0.445^{+0.080}_{-0.082}$ & $0.295^{+0.006}_{-0.006}$ & $0.79^{+0.07}_{-0.07}$ & $1.09^{+0.02}_{-0.02}$ & --- & ---&  $1.23/87$ \\
            
            25/01/16 & $2.033^{+0.036}_{-0.035}$ & $0.449^{+0.089}_{-0.092}$ & $0.299^{+0.007}_{-0.007}$ & $0.92^{+0.09}_{-0.09}$ & $1.08^{+0.02}_{-0.02}$ & --- & --- & $0.83/79$ \\
            
            25/01/29 & $2.183^{+0.033}_{-0.034}$ & $0.387^{+0.087}_{-0.085}$ & $0.245^{+0.005}_{-0.005}$ & $0.58^{+0.06}_{-0.06}$ & $0.93^{+0.03}_{-0.03}$ & --- & --- & $1.08/87$ \\
            
            25/01/31 & $2.071^{+0.026}_{-0.026}$ & $0.559^{+0.066}_{-0.067}$ & $0.436^{+0.007}_{-0.007}$ & $0.86^{+0.05}_{-0.05}$ & $1.67^{+0.02}_{-0.02}$ & --- & --- & $0.99/142$ \\
            
            25/02/04 & $2.154^{+0.027}_{-0.026}$ & $0.320^{+0.066}_{-0.067}$ & $0.401^{+0.007}_{-0.007}$ & $0.57^{+0.06}_{-0.06}$ & $1.48^{+0.03}_{-0.03}$ & --- & --- & $0.89/123$ \\
            
            25/02/12 & $1.922^{+0.029}_{-0.028}$ & $0.497^{+0.068}_{-0.069}$ & $0.391^{+0.007}_{-0.007}$ & $1.20^{+0.07}_{-0.08}$ & $1.39^{+0.02}_{-0.02}$ & --- & --- & $1.04/116$ \\
            
            25/02/14 & $1.941^{+0.032}_{-0.031}$ & $0.447^{+0.073}_{-0.075}$ & $0.459^{+0.009}_{-0.009}$ &$1.16^{+0.09}_{-0.09}$ & $1.60^{+0.03}_{-0.03}$ & --- & --- &  $0.88/101$\\
            
            25/02/17 & $1.933^{+0.025}_{-0.025}$ & $0.456^{+0.058}_{-0.059}$ & $0.553^{+0.009}_{-0.009}$ & $1.18^{+0.07}_{-0.07}$ & $1.93^{+0.03}_{-0.03}$ & --- & --- & $1.24/162$ \\
            
            25/02/20 & $1.910^{+0.026}_{-0.026}$ & $0.451^{+0.058}_{-0.059}$ & $0.467^{+0.008}_{-0.008}$ & $1.26^{+0.08}_{-0.08}$ & $1.62^{+0.02}_{-0.02}$ & --- & --- & $0.88/152$ \\
            
            25/02/23 & $2.045^{+0.029}_{-0.028}$ & $0.459^{+0.072}_{-0.073}$ & $0.368^{+0.007}_{-0.007}$ & $0.89^{+0.07}_{-0.07}$ & $1.34^{+0.02}_{-0.02}$ & --- & --- & $1.16/112$ \\
            
            25/02/26 & $2.247^{+0.037}_{-0.036}$ & $0.597^{+0.111}_{-0.114}$ & $0.248^{+0.006}_{-0.006}$ & $0.62^{+0.05}_{-0.05}$ & $1.05^{+0.03}_{-0.03}$ & --- & --- & $1.09/70$ \\
            
            25/03/01 & $1.971^{+0.032}_{-0.031}$ & $0.277^{+0.070}_{-0.071}$ & $0.341^{+0.007}_{-0.007}$ & $1.13^{+0.14}_{-0.14}$ & $1.11^{+0.02}_{-0.02}$ & --- & --- & $1.19/95$ \\ 
         \hline
        \end{tabular}
     }
     \tablefoot{\textit{Column (1):} Date of the observations. \textit{Column (2):} Spectral index at 1~keV. \textit{Column (3):} Curvature parameter. \textit{Column (4):} 0.3-10~keV unabsorbed flux in unit of $10^{-9}$ erg cm$^{-2}$ s$^{-1}$. \textit{Column (5):} Energy of the synchrotron peak. \textit{Column (6):} Flux at the synchrotron peak in unit of $10^{-10}$ erg cm$^{-2}$ s$^{-1}$. \textit{Column (7):} Calibration constant for SVOM/MXT. \textit{Column (8):} Calibration constant for SVOM/ECLAIRs. \textit{Column (9):} Reduced $\chi^2$ with the associated degree of freedom.}
     \label{tab:spectral_parameters}
 \end{table*}

\section{Search for hysteresis in the $\mathrm{HR}-F_{0.3-10\,\mathrm{keV}}$ plane}\label{app:hysteresis}

To quantify the presence of loops in the $\mathrm{HR}-F_{0.3-10\,\mathrm{keV}}$ plane, we computed the area enclosed by successive time-ordered observations. The area was calculated using the Gauss (shoelace) formula \citep{Braden1986TheSA}. Each loop was required to contain at least three points. For the $i$-th loop, defined by observations with indices ranging from $k_{\mathrm{min}}$ to $k_{\mathrm{max}}$, the enclosed area is given by

\begin{equation}
    a_i = \frac{1}{2} \sum_{k=k_{\mathrm{min}}}^{k_{\mathrm{max}}} (F_{0.3-10\,\mathrm{keV},k}\cdot HR_{k+1}- F_{0.3-10\,\mathrm{keV},k+1}\cdot HR_{k}).
    \label{eq:gauss_area}
\end{equation}

With this convention, negative (positive) areas correspond to CW (CCW) rotations. Applying this method to all possible sequences yielded 435 candidate loops. To assess their statistical significance, we performed 10,000 bootstrap simulations by randomizing the observation indices. The resulting distributions of simulated loop areas showed that none of the observed loops exceed a significance of $3\sigma$.

\end{appendix} 

\end{document}